\documentclass[cameraready]{Interspeech}
\usepackage{booktabs}
\usepackage{multirow}
\usepackage{graphicx}
\usepackage{xcolor}
\usepackage{amsmath,amssymb}
\usepackage{comment}
\usepackage{siunitx}
\usepackage{placeins}
\usepackage{cite}
\usepackage{makecell}
\usepackage{subcaption}
\usepackage{svg}

\newcommand{\indicator}{\mathbb{I}}
\newcommand{\Qset}{\mathcal{Q}}
\newcommand{\acc}{\mathrm{Acc}}

\newcommand{\nmodels}{8}

\title{All That Glitters Is Not Audio: Rethinking Text Priors and Audio Reliance in Audio-Language Evaluation}

\author[affiliation={1}, equalcontribution]{Leonardo Haw-Yang Foo}{}
\author[affiliation={1}, equalcontribution]{Chih-Kai Yang}{}
\author[affiliation={1}]{Chen-An Li}{}
\author[affiliation={1}]{Ke-Han Lu}{}
\author[affiliation={1,2}]{Hung-yi Lee}{}

\address{
  $^1$National Taiwan University, $^2$NTU Artificial Intelligence Center of Research Excellence (NTU AI-CoRE)
}

\email{leonardofoohy@gmail.com, chihkaiyang1124@gmail.com, hungyilee@ntu.edu.tw}


\begin{document}
\maketitle

\begin{abstract}
Large Audio-Language Models show consistent performance gains across speech and audio benchmarks, yet high scores may not reflect true auditory perception. 
If a model can answer questions without processing the acoustic signal, the benchmark fails as a measure of auditory understanding. 
We present a diagnostic framework using two axes: text prior, which measures answerability from text and general knowledge alone, and audio reliance, which assesses actual dependency on the acoustic signal. 
Evaluating eight LALMs across three benchmarks, we find that models retain 60–72\% of their full audio scores even without any audio input.
Moreover, among items that require audio, only 3.0–4.2\% need the complete audio clip; the majority can be resolved using localized fragments. 
These findings challenge the assumption that benchmark performance equals robust audio understanding, and we conclude with practical guidelines for improving evaluation reliability and benchmark design.
\end{abstract}

\section{Introduction}
\label{sec:intro}

The rapid development of Large Audio-Language Models (LALMs)~\cite{qwen2audio, qwen25omni, qwen3omni, af3, phi4multimodal, desta25audio, voxtral, tp1, copilot, desta, desta2, lin2025preliminary, audioflamingo, tang2024salmonn, wang-etal-2024-blsp}, which extend Large Language Models (LLMs)~\cite{grattafiori2024llama, yang2025qwen3} with auditory perception and knowledge~\cite{yang2025audiolens, sake}, has led to consistent performance gains on speech and audio benchmarks~\cite{mmau, mmar, mmaupro, audiobench, air-bench, dynamic-superb, dynamic-superb-2, sakura, listenfairly, speechifeval, yang2026mugen, yang2025speechr, voxeval}. These improvements are often interpreted as evidence of strong auditory understanding~\cite{arora2025on, audio-eval-survey}.

However, benchmark performance may also be influenced by \emph{text prior}, where questions can be answered from textual cues or general knowledge without processing the audio signal. Similar effects have been observed before. In natural language inference, hypothesis-only baselines~\cite{poliak-etal-2018-hypothesis} achieve high accuracy without the premise, exposing annotation artifacts and lexical shortcuts~\cite{gururangan2018annotation}. Likewise, question-only baselines in visual question answering show that strong language priors can enable correct answers without grounding in visual evidence~\cite{goyal2017vqav2, chen2024mmstar}. Audio-language benchmarks may exhibit a similar issue. Table~\ref{tab:text_prior_example} shows a case solvable from linguistic knowledge alone without the audio. In such cases, benchmark scores may reflect text-based reasoning rather than auditory perception.

\begin{figure}[hbtp]
    \centering
    \includegraphics[width=\linewidth]{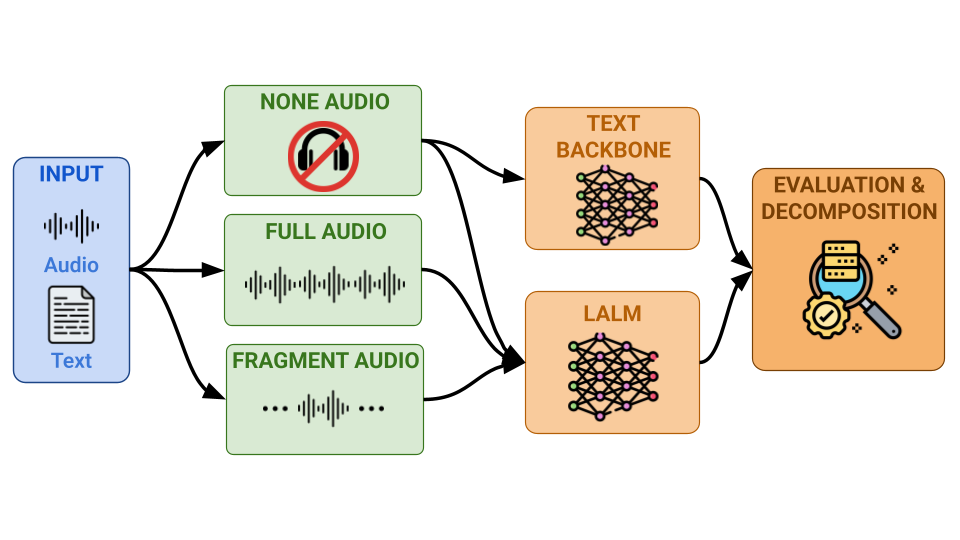}
    \vspace{-30pt}
    \caption{Overview of the proposed diagnostic framework.}
    \label{fig:overview}
    \vspace{-10pt}
\end{figure}


Recent studies attempt to verify whether models use audio by modifying the input signal and observing performance changes~\cite{silence-matters, mmau, mmaupro}. A related approach replaces the audio with a silent signal such as 30 seconds of silence to proxy text-only behavior~\cite{he2025audiomcq}. However, silence itself has been shown to interfere with model outputs independently of task content~\cite{silence-matters}, making it an imperfect substitute for a true text-only condition. Consequently, these studies do not quantify how much benchmark performance can be achieved without audio.

In this work, we analyze benchmark behavior along two diagnostic axes.
\textbf{Text Prior} measures how much of a benchmark can be solved from the textual prompt and general knowledge alone.
\textbf{Audio Reliance} measures how much model performance actually depends on the acoustic signal.
To analyze audio reliance, we evaluate not only the standard full-audio condition but also settings where the audio is partitioned into multiple equal-duration segments that are evaluated independently. By observing how performance changes when only partial audio is available, we can determine whether a benchmark requires global audio understanding or merely short local cues.

We apply this framework to three benchmarks, MMAU~\cite{mmau}, MMAR~\cite{mmar}, and MMAU-Pro~\cite{mmaupro}, evaluating eight advanced LALMs. Our results reveal a substantial grounding gap: even without audio input, models retain 60–72\% of their full-audio accuracy, indicating that a large portion of benchmark performance can be reproduced using text-only information. Furthermore, among items that require audio for correct prediction, only 3.0–4.2\% depend on information distributed across the entire audio clip, while most can already be solved from a single localized fragment.

These findings suggest that current benchmarks often measure a mixture of textual priors and localized audio cues rather than consistently requiring holistic audio understanding. To facilitate more reliable evaluation, we introduce a diagnostic framework with three main components:
\begin{itemize}
\item Two diagnostic axes, text prior and audio reliance, to quantify dependence on textual versus acoustic signals.
\item A large-scale decomposition across eight LALMs and three benchmarks showing that a substantial portion of benchmark scores can be achieved without audio input.
\item A temporal granularity analysis revealing that most audio-dependent items can be solved from short audio fragments rather than requiring the full clip.
\end{itemize}

Together, these analyses provide a clearer picture of how current benchmarks utilize audio information and offer practical guidance for designing evaluations that more faithfully measure auditory understanding in LALMs.
\section{Methodology}
\label{sec:method}

We propose two diagnostic axes for auditing audio-language benchmarks.
\textit{Text prior} (\S\ref{sec:text_prior}) quantifies how much
performance is achievable without audio.
\textit{Audio reliance} (\S\ref{sec:audio_reliance}) measures how much
of the audio signal models actually use.

\subsection{Text Prior}
\label{sec:text_prior}

A desirable audio-language benchmark should require access to the auditory signal to be solved correctly. 
Ideally, performance should approach random when audio is removed. 
In practice, however, models may exploit biases in the textual components of a benchmark, such as question phrasing or answer options, and achieve non-trivial accuracy without processing audio. 
We refer to this phenomenon as \emph{text prior}, defined as the extent to which a benchmark can be solved using textual information alone. 
When text prior is strong, benchmark scores may substantially overestimate a model’s auditory capability, as correct answers can be obtained without attending to the audio.

To quantify text prior, we consider three controlled settings:
\begin{enumerate}
    \item \textbf{Text Backbone (TB):} the original text-only backbone of the LALM, prior to multimodal training, is evaluated using only the textual question. 
    This measures how much of the benchmark can be solved by a purely text-based model.

    \item \textbf{None:} the LALM is evaluated without audio input and receives only the textual question. This reflects the model's text-only behavior after multimodal adaptation. Unlike prior work that proxies text-only behavior by replacing audio with silence~\cite{he2025audiomcq}, we omit the audio input entirely to avoid potential confounds introduced by silent audio~\cite{silence-matters}.
    
    \item \textbf{Full:} the standard setting in which both audio and text are provided to the LALM.
\end{enumerate}

Comparing these settings reveals how performance changes across conditions. 
High accuracy under TB indicates strong text prior. 
The TB–None difference reflects how multimodal training alters language-only behavior, while the None–Full gap shows the additional effect of the audio signal.

To summarize the strength of text prior after multimodal training, we define the \emph{text-prior rate}:
$R_{\mathrm{TP}} = \acc_{\mathrm{none}} / \acc_{\mathrm{full}}$,
where $\acc_{\mathrm{none}}$ and $\acc_{\mathrm{full}}$ denote the model accuracies under the None and Full settings, respectively. 
This quantity represents the fraction of full-audio accuracy that the LALM can achieve without access to audio.

\subsection{Audio Reliance}
\label{sec:audio_reliance}

Beyond text prior, we analyze how much audio is actually required for correct prediction. 
Even when Full-audio evaluation improves over text-only settings, it remains unclear whether models rely on the entire audio or only local cues. 
If small fragments already preserve high accuracy, the benchmark may not require global understanding, suggesting that models rely on short, localized signals correlated with the correct answer.

To examine this, we partition each audio clip into $N$ equal-duration contiguous segments and evaluate each segment independently using the same textual prompt.
For each item $q \in \Qset$ and segment index $k$, let $c_{q,k}^{(N)} \in \{0,1\}$ denote whether the prediction based on the $k$-th segment is correct. We first compute the average segment accuracy across its $N$ segments, and then average this quantity over all items in $\Qset$.
Finally, we normalize by the full-audio accuracy $\acc_{\mathrm{full}}$ to obtain the \emph{retention rate} $R_N$, which measures how much performance is preserved when only $1/N$ of the audio is provided:

\begin{equation}
\begin{aligned}
  R_N
  &= \frac{1}{|\Qset| \cdot \acc_{\mathrm{full}}}
     \sum_{q \in \Qset}
     \frac{1}{N}
     \sum_{k=1}^{N}
     \indicator[c_{q,k}^{(N)} = 1].
\end{aligned}
\label{eq:retention}
\end{equation}

The retention rate measures how much performance is preserved when only partial audio is available.
Values near 1 indicate that fragments are nearly as informative as the full audio, suggesting reliance on local cues rather than global context.
A significantly lower value indicates substantial degradation without full context, implying greater use of global information.

By varying $N$, we can analyze how performance changes as temporal granularity increases, thereby characterizing the degree to which the benchmark requires full-audio context.

\begin{table}[t]
    \centering
    \vspace{-10pt}
    \caption{Item categorization by correctness under Full, None, and fragment conditions.
    \checkmark and $\times$ represent that the item is correctly or incorrectly answered under the specified conditions.}
    \label{tab:categories}
    \small
    \setlength{\tabcolsep}{4pt}
    \begin{tabular}{@{}lccc@{}}
        \toprule
        Category & Full & None & Fragment \\
        \midrule
        Text-Solvable (TS)       & \checkmark & \checkmark & ---               \\
        Audio-Needed (AN)        & \checkmark & $\times$   & ---               \\
        Fragment-Sufficient (FS) & \checkmark & $\times$   & $\exists$ correct \\
        Cross-Segment (XS)       & \checkmark & $\times$   & all incorrect     \\
        Audio-Harmful (AH)       & $\times$   & \checkmark & ---               \\
        Unsolvable (UN)          & $\times$   & $\times$   & ---               \\
        \bottomrule
    \end{tabular}
\end{table}
\subsection{Score Decomposition}
\label{sec:decomposition}

To understand the interplay between textual priors and audio reliance, we perform a joint analysis at the item level. 
By comparing model correctness across the Full, None, and Fragment audio conditions, we classify each item into one of five mutually exclusive categories (Table~\ref{tab:categories}):
\begin{itemize}
    \item \textbf{Text-Solvable (TS)}: The model answers correctly regardless of whether audio is provided.
    \item \textbf{Audio-Needed (AN)}: Audio is required for a correct answer.
    \item \textbf{Fragment-Sufficient (FS)}: Audio is required, but a fragment already provides sufficient information.
    \item \textbf{Cross-Segment (XS)}: The full audio is required; no single fragment alone is sufficient. Note that AN = FS $\cup$ XS.
    \item \textbf{Audio-Harmful (AH)}: The model answers correctly without audio but fails when audio is provided.
    \item \textbf{Unsolvable (UN)}: The model fails regardless of whether audio is available.
\end{itemize}

This decomposition separates text-solvable items from audio-dependent ones and further distinguishes fragment-sufficient cases from those requiring full audio context. 
The five categories are mutually exclusive and exhaustive, so within each benchmark we have $|\mathrm{TS}|+|\mathrm{FS}|+|\mathrm{XS}|+|\mathrm{AH}|+|\mathrm{UN}|=|\Qset|.$
\section{Experimental Setups}
\label{sec:setups}
\begin{table}[t]
    \centering
    \caption{Evaluated LALMs and their text backbones.}
    \label{tab:models}
    \vspace{-5pt}
    \footnotesize
    \setlength{\tabcolsep}{3pt}
    \begin{tabular}{@{}ll@{}}
        \toprule
        LALM & Text Backbone \\
        \midrule
        Audio-Flamingo-3~\cite{af3}   & Qwen2.5-7B-Instruct~\cite{qwen2024qwen25} \\
        DeSTA-2.5~\cite{desta25audio}          & Llama-3.1-8B-Instruct~\cite{grattafiori2024llama} \\
        Phi-4-Multimodal~\cite{phi4multimodal}   & Phi-4-Mini-Instruct~\cite{abouelenin2025phi} \\
        Qwen2-Audio-7B~\cite{qwen2audio}     & Qwen-7B-Chat~\cite{qwen1} \\
        Qwen2.5-Omni-7B~\cite{qwen25omni}   & Qwen2.5-7B-Instruct~\cite{qwen2024qwen25} \\
        Qwen3-Omni (I)~\cite{qwen3omni}    & Qwen3-30B-A3B-Instruct~\cite{yang2025qwen3} \\
        Qwen3-Omni (T)~\cite{qwen3omni}    & Qwen3-30B-A3B-Thinking~\cite{yang2025qwen3} \\
        Voxtral-Mini-3B~\cite{voxtral}    & Ministral-3B~\cite{ministral} \\
        \bottomrule
    \end{tabular}
    \vspace{-15pt}
\end{table}
\begin{table*}[htbp]
    \centering
    \caption{Analysis for text prior. 
    (a) Full, None, and Text Backbone (TB) results with text-prior rate $R_{\mathrm{TP}}=\mathrm{None}/\mathrm{Full}$ (\S\ref{sec:text_prior}). 
    MMAU-Pro includes MCQ items only. 
    $^\dagger$ MoE; 3B active parameters. 
    (b) Examples of textual priors in MMAU.}
    \label{tab:text_prior_all}
    \vspace{-5pt}
    \begin{subtable}[htbp!]{0.68\textwidth}
         \centering
         \caption{Full, None, TB, and $R_{\mathrm{TP}}$ across three benchmarks.}
         \label{tab:text_prior}
         \vspace{-3pt}
         \resizebox{\textwidth}{!}{
         \begin{tabular}{@{}l l cccc cccc cccc@{}}
            \toprule
            & & \multicolumn{4}{c}{MMAU} & \multicolumn{4}{c}{MMAR} & \multicolumn{4}{c}{MMAU-Pro} \\
            \cmidrule(lr){3-6} \cmidrule(lr){7-10} \cmidrule(lr){11-14}
            Model & Size & Full & None & TB & $R_{\mathrm{TP}}$ & Full & None & TB & $R_{\mathrm{TP}}$ & Full & None & TB & $R_{\mathrm{TP}}$ \\
            \midrule
            Audio-Flamingo-3   & 8.4B  & 75.0 & \textbf{60.9} & 45.5 & \textbf{81.2} & 58.8 & 33.1 & 35.3 & 56.3 & 52.7 & \textbf{44.1} & 31.2 & \textbf{83.7} \\
            DeSTA-2.5          & 8.8B  & 65.2 & 28.1 & 28.4 & 43.1 & 46.4 & 26.1 & 26.2 & 56.2 & 43.5 & 31.3 & 20.3 & 72.0 \\
            Phi-4-Multimodal   & 5.6B  & 60.4 & 29.0 & 28.9 & 48.0 & 46.1 & 27.6 & 28.3 & 59.9 & 43.7 & 28.6 & 29.9 & 65.5 \\
            Qwen2-Audio-7B     & 8.2B  & 63.9 & 38.3 & 38.5 & 59.9 & 46.3 & 26.0 & 22.5 & 56.2 & 44.8 & 31.4 & 28.2 & 70.1 \\
            Qwen2.5-Omni-7B    & 10.7B & 74.8 & 48.7 & 45.5 & 65.1 & 63.9 & 41.3 & 35.3 & 64.6 & 57.7 & 39.3 & 31.2 & 68.2 \\
            Qwen3-Omni (I)     & 30B$^\dagger$  & \textbf{77.4} & 56.6 & \textbf{50.8} & 73.1 & \textbf{69.7} & \textbf{44.1} & \textbf{37.6} & 63.3 & \textbf{59.5} & 43.2 & \textbf{41.0} & 72.6 \\
            Qwen3-Omni (T)     & 30B$^\dagger$  & 76.2 & 55.8 & 38.6 & 73.2 & 70.3 & 41.9 & 31.6 & 59.6 & 56.5 & 40.5 & 33.8 & 71.7 \\
            Voxtral-Mini-3B    & 4.7B  & 55.9 & 39.6 & 23.0 & 70.8 & 50.9 & 33.8 & 26.3 & 66.4 & 41.7 & 30.0 & 20.0 & 71.9 \\
            \midrule
            \textbf{OVERALL}   & -- & 68.6 & 44.6 & 37.4 & 65.1 & 56.5 & 34.2 & 30.4 & 60.5 & 50.0 & 36.0 & 29.5 & 72.1 \\
            \textit{Chance Level}    & -- & 25.0 & 25.0 & 25.0 & --    & 25.0 & 25.0 & 25.0 & --    & 25.9 & 25.9 & 25.9 & -- \\
            \bottomrule
        \end{tabular}
        }
    \end{subtable}
    \begin{subtable}[htbp!]{0.305\textwidth}
        \centering
        \caption{Example of text prior in MMAU.}
        \label{tab:text_prior_example}
        \vspace{-3pt}
        \resizebox{\textwidth}{!}{
        \begin{tabular}{p{0.25\textwidth} p{0.70\textwidth}}
            \toprule
            \textbf{Field} & \textbf{Content} \\
            \midrule
            Example &
            Based on the given audio, identify the source of the moo sound. \\
            \midrule
            Explanation &
            The onomatopoeic word ``moo'' strongly implies the sound source is a cow, allowing a text-only LLM to answer correctly without listening to the audio. \\
            \bottomrule
        \end{tabular}
        }
    \end{subtable}
    \vspace{-10pt}
\end{table*}

\subsection{Benchmarks}
\label{sec:benchmarks}

We evaluate on three public audio-language benchmarks. 
\textbf{MMAU}~\cite{mmau} is a 10,000-item MCQ dataset covering sound, music, and speech; we use the 1,000-item test-mini split. 
\textbf{MMAR}~\cite{mmar} contains 1,000 MCQ items organized by four cognitive layers and seven modality combinations. 
\textbf{MMAU-Pro}~\cite{mmaupro} includes 5,305 items spanning MCQ, open-ended QA, and audio instruction following across 12 categories.

\subsection{Models}
\label{sec:models}

We analyze \nmodels{} LALMs ranging from 3B to 30B parameters, each paired with its corresponding text backbone (Table~\ref{tab:models}). 
Qwen3-Omni is evaluated in both \emph{Instruct} and \emph{Thinking} modes, with the latter enabling inference-time reasoning. 
All models use greedy decoding except Qwen3-Omni Thinking, which follows the recommended temperature of 0.6.

\subsection{Evaluation Protocols}
\label{sec:protocols}

Official string-match scorers penalize verbose but correct outputs;
on 185 manually annotated format-sensitive cases from MMAU and MMAR,
the LLM judge matches human annotations in 97.2\% of cases versus
26.0\% for string-match.
We therefore adopt a hybrid MCQ scorer: regular expression answer extraction
followed by Claude 4.5 Haiku~\cite{claude45haiku} at
temperature\,=\,0 for cases where regex fails.
All MCQ results use this scorer unless otherwise noted.
For MMAU-Pro open-ended and instruction-following items, we retain the
benchmark's original evaluation: a Qwen2.5-7B-Instruct LLM judge for
open-ended items and rule-based format checking for
instruction-following.
For the audio reliance analysis (\S\ref{sec:audio_reliance}), we
evaluate $N \in \{2,3,4,5\}$ equal-duration fragments per clip.
\section{Results}
\label{sec:results}

\subsection{Results on Text Prior}
\label{sec:results_text_prior}
Table~\ref{tab:text_prior} reports the accuracies under the Full, None, and Text Backbone (TB) settings for all models and benchmarks, together with the corresponding text-prior rate $R_{\mathrm{TP}}$.

TB accuracy substantially exceeds chance for most models across benchmarks.
Averaged across models, TB surpasses chance by 12.4\%, 5.4\%, and 3.6\% on MMAU, MMAR, and MMAU-Pro, respectively.
Qwen-family text backbones show particularly high TB accuracy, with Qwen3-30B-A3B-Instruct reaching 50.8\%, 37.6\%, and 41.0\%.
This gap between TB and chance indicates that many items can be solved using textual cues alone, revealing strong text priors under modern LLM backbones commonly used in LALM training.
Consequently, benchmark scores may partly reflect text-based reasoning rather than genuine auditory understanding.
Table~\ref{tab:text_prior_example} presents an MMAU example solvable from text alone.

We compare accuracy under the Full and None conditions.
For most models, None accuracy exceeds TB, suggesting that multimodal training strengthens the text prior of LALMs beyond that of their original LLM backbones.
Using the text-prior rate $R_{\mathrm{TP}}$ to compare None and Full, most models retain over 60\% of their Full accuracy without audio.
The effect is particularly pronounced in some cases, such as Audio-Flamingo-3 on MMAU and MMAU-Pro, and the average $R_{\mathrm{TP}}$ across models also exceeds 60\%.
These results indicate that benchmark performance can be substantially driven by text prior.

Overall, our findings reveal a significant text prior issue that challenges the common assumption that these benchmarks primarily evaluate auditory understanding in LALMs.
This issue has largely been overlooked and warrants careful consideration in future benchmark design and evaluation.

\begin{figure}[t]
    \centering
    \includegraphics[width=0.8\linewidth]{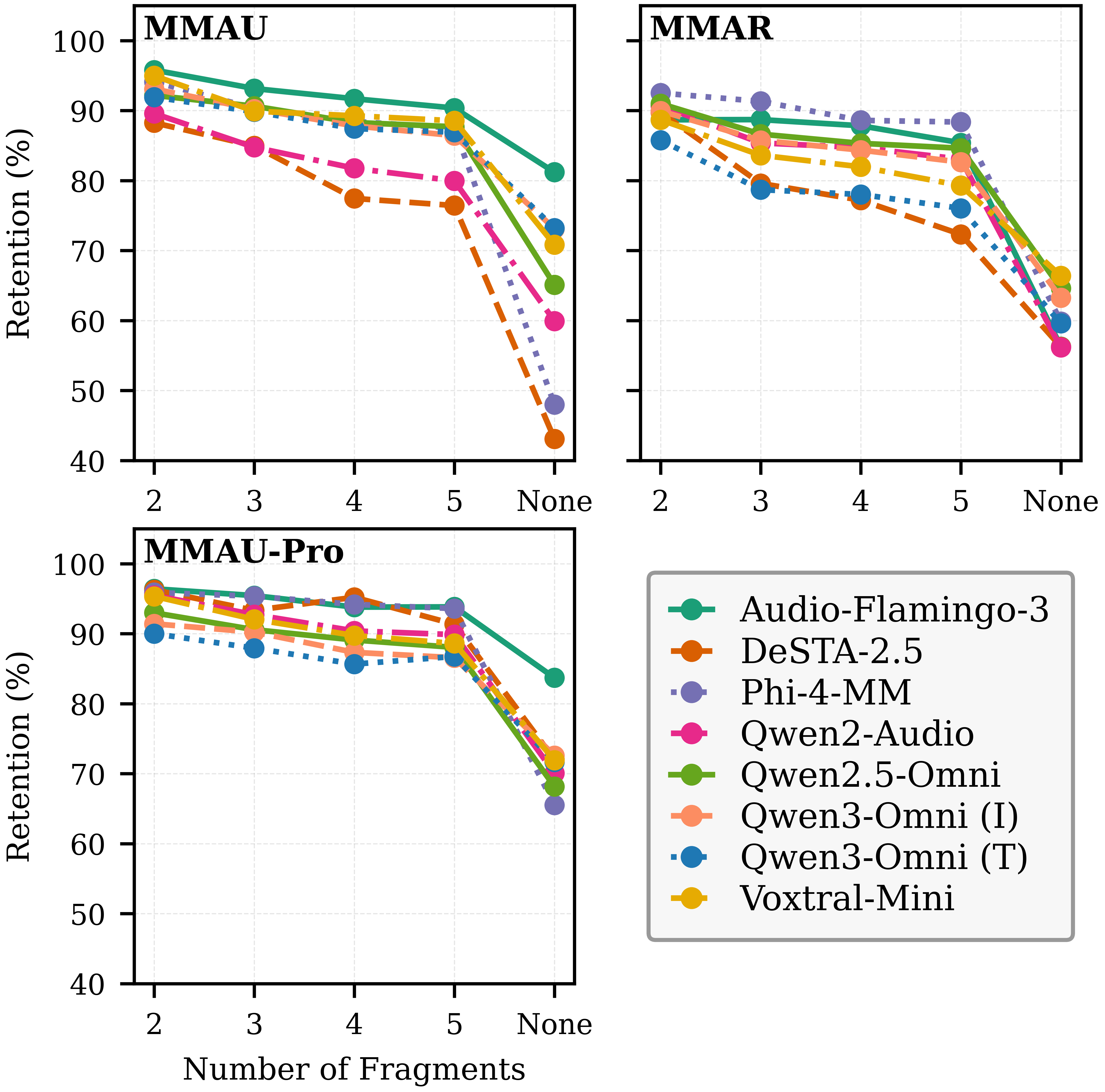}
    \caption{Retention rate (\%) across three benchmarks for eight models. Higher retention indicates greater reliance on information preserved in short audio fragments.}
    \label{fig:retention_trends}
\end{figure}

\begin{figure}[t]
    \centering
    \includegraphics[width=1.0\linewidth]{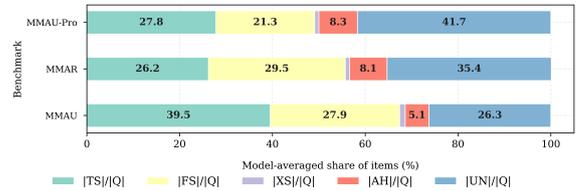}
    \caption{Model-averaged stacked distribution of item categories across the three benchmarks.}
    \label{fig:audio_reliance_breakdown}
    \vspace{-5pt}
\end{figure}

\subsection{Results on Audio Reliance}
\label{sec:results_audio_reliance}

For audio reliance, we examine not only the performance gap between the Full and None settings, but also how model accuracy changes when only partial audio is available.
Specifically, we combine two complementary analyses: (i) the retention curve $R_N$ (Eq.~\ref{eq:retention}), which measures how performance changes when the audio is reduced to $1/N$ of the original clip, and (ii) the per-item decomposition, which categorizes items according to how audio contributes to the prediction.

Figure~\ref{fig:retention_trends} shows the retention curves on MMAU, MMAR, and MMAU-Pro.
As the number of fragments per clip increases ($N{=}2$ to $5$), retention gradually declines, indicating performance degradation with reduced audio.
However, retention remains relatively high, suggesting that much of the required information is preserved even in short fragments.
Together with the earlier observation that Full significantly outperforms None, this implies that although audio improves over text-only input, much of the gain can be recovered from short audio fragments.

To better understand this behavior at the item level, we analyze the category decomposition defined in \S\ref{sec:decomposition}.
Figure~\ref{fig:audio_reliance_breakdown} reports the distribution of items across the five categories, averaged over \nmodels{} models.
A substantial portion of items are \emph{Text-Solvable} (TS), accounting for 26.2–39.5\% across the three benchmarks, confirming the strong presence of text prior.
Meanwhile, items that genuinely require audio (
\(|\mathrm{AN}|=|\mathrm{FS}|+|\mathrm{XS}|\) account for 22.2–30.4\% of the datasets.

Among the audio-needed items, the majority are \emph{Fragment-Sufficient} (FS).
The \emph{Cross-Segment} (XS) category, which indicates that no single fragment suffices, remains rare.
As shown in Table~\ref{tab:audio_needed}, XS accounts for only 3.0–4.2\% of audio-needed items on average, meaning that nearly 96\% of audio-dependent items can already be solved from at least one local fragment.

Taken together, these results suggest that while audio does contribute to benchmark performance, much of the useful information is localized rather than distributed across the entire audio clip.
In other words, current benchmarks contain relatively few items where cross-segment cues are indispensable, and many audio-dependent cases can already be resolved using short segments of the signal.
These findings suggest that current benchmarks measure a mixture of text priors and localized audio cues, rather than robust holistic audio understanding.

\subsection{Fine-grained Analysis}
\label{sec:fine_grained}

Table~\ref{tab:domain} analyzes audio reliance across task categories using the Full–None gap as a proxy for audio dependency.
Instruction-following (IF) tasks show the largest drop (36.9\%) but account for only 1.6\% of MMAU-Pro items.
Among multiple-choice domains, speech is the most audio-dependent, with gaps of 24.7–27.7\%.
In contrast, sound and music retain over 93\% of Full accuracy at $N{=}2$, indicating that short fragments often suffice.
Open-ended items in MMAU-Pro even perform slightly better without audio, suggesting that audio can sometimes act as a distractor when strong text priors exist.
Overall, these results indicate that while certain tasks depend on audio, much of the benchmark performance can still be explained by text prior and localized audio cues.

\begin{table}[t]
    \centering
    \caption{Analysis of audio dependency. 
    \textbf{Audio-Needed (AN)} represents the proportion of items requiring audio ($|\mathrm{AN}|/|\Qset|$). 
    \textbf{Global Dependency} represents the proportion of AN items that require full-context audio ($|\mathrm{XS}|/|\mathrm{AN}|$).
    Results are averaged across \nmodels{} models.}
    \vspace{-5pt}
    \label{tab:audio_needed}
    \small
    \setlength{\tabcolsep}{5pt}
    \renewcommand{\arraystretch}{0.8}
    \resizebox{0.98\linewidth}{!}{
    \begin{tabular}{@{}l c c c@{}}
        \toprule
        & Audio-Needed & \multicolumn{2}{c}{Global Dependency} \\
        \cmidrule(lr){2-2} \cmidrule(l){3-4}
        Benchmark & Mean (\%) & Mean (\%) & Range (Min--Max) \\
        \midrule
        MMAU     & 29.1 & 4.2 & 2.2--5.7 \\
        MMAR     & 30.4 & 3.0 & 1.5--5.4 \\
        MMAU-Pro & 22.2 & 4.0 & 2.1--8.0 \\
        \bottomrule
    \end{tabular}
    }
\end{table}

\begin{table}[t]
    \centering
    \caption{Model-averaged accuracy (\%) by audio category, ordered by audio dependency (Full–None gap).
BM: benchmark (Pro = MMAU-Pro).
$N{=}2$: average over two equal-duration fragments.
Results average over \nmodels{} models.
MMAU-Pro instruction-following (IF) and Open use the original evaluation; others use the LLM judge (\S\ref{sec:protocols}).}
    \vspace{-5pt}
    \label{tab:domain}
    \small
    \setlength{\tabcolsep}{4.5pt}
    \resizebox{0.98\linewidth}{!}{
    \begin{tabular}{@{}ll r rrrr@{}}
        \toprule
        Category & BM & \#Items & Full & $N{=}2$ & None & F$-$N (\%) \\
        \midrule
        IF       & Pro  &    87 & 52.6 & 36.2 & 15.7 & 36.9 \\
        Speech   & MMAU &   333 & 67.2 & 57.5 & 39.5 & 27.7 \\
        Speech   & MMAR &   294 & 61.9 & 52.7 & 34.9 & 27.0 \\
        Sound    & MMAU &   333 & 72.8 & 68.3 & 47.2 & 25.6 \\
        Speech   & Pro  &   891 & 57.6 & 49.6 & 33.0 & 24.7 \\
        Music    & MMAU &   334 & 65.9 & 64.6 & 47.1 & 18.8 \\
        Sound    & Pro  & 1{,}047 & 44.5 & 43.7 & 41.6 & 2.9 \\
        Open     & Pro  &   625 & 66.0 & 65.6 & 67.9 & $-$1.9 \\
        \bottomrule
    \end{tabular}
}

\vspace{-5mm}
\end{table}

\section{Recommended Practices}

We hope this study provides practical guidance for LALM research. Benchmark designers should measure \emph{text prior} to ensure tasks cannot be solved using text-only cues and thus genuinely assess auditory understanding. Model developers should also compare performance with and without audio to verify that improvements arise from auditory reasoning rather than textual shortcuts. In addition, benchmark designers can measure \emph{audio reliance} using the retention rate to quantify dependence on the audio signal. Tasks targeting holistic long-form understanding should exhibit stronger audio reliance, while tasks based on local cues may tolerate higher retention under partial audio. Together, text prior and audio reliance provide practical indicators for whether a benchmark aligns with its intended objective.

\section{Conclusion}
\label{sec:conclusion}
We analyze how current audio-language benchmarks depend on audio by decomposing performance into text prior and audio reliance. Across three benchmarks and eight LALMs, models retain 60–72\% of their full accuracy without audio, revealing strong text priors. Under partial audio, only 3.0–4.2\% of items require cross-segment information, while most can be solved from short segments. These findings suggest that benchmark performance often reflects a mixture of text priors and localized audio cues. To improve evaluation reliability, we advocate reporting text-prior baselines and analyzing audio reliance to ensure benchmarks measure their intended abilities.

\ifcameraready
\section{Acknowledgments}

We acknowledge the computational and storage support provided by the National Center for High-performance Computing (NCHC) of the National Applied Research Laboratories (NARLabs) in Taiwan. This work was supported by the Ministry of Education (MOE) of Taiwan under the project Taiwan Centers of Excellence in Artificial Intelligence, through the NTU Artificial Intelligence Center of Research Excellence (NTU AI-CoRE)
\fi

\section{Generative AI Use Disclosure}

Generative AI tools were used in this paper solely for language polishing and writing refinement. In addition, large language models were utilized as judges for the automatic evaluation.

\bibliographystyle{IEEEtran}
\bibliography{references}

\end{document}